# Effects of Antimony Deposition on Field-Emission Current Density of Ge/Si

Veronika Burobina

*Ioffe Institute, 26 Politekhnicheskaya, St Petersburg 194021, Russia*

*Electronic Mail: nikabour@mail.ioffe.ru*

To estimate the field-emission current density of a Ge/Si heterosystem, 20-nm germanium/ silicon (100) samples were grown by molecular beam epitaxy. The surface of one sample was covered with a layer of antimony, which was removed in vacuum prior to the samples being measured. A second sample of Ge/Si was exposed to room air in the absence of antimony. The current-voltage characteristics of both samples obtained by scanning tunneling microscopy were discovered to be in agreement with classical Fowler-Nordheim theory. The density of emission current from Ge nanocrystal exceeds the density of emission current from the wetting layer of Ge/Si. The density of emission current of pure Ge nanocrystal is less than the density of emission current of Ge nanocrystal with adsorption layers.

## INTRODUCTION

Field emission [1], [2] is a quantum effect in which the emission of electrons from a cathode does not require expenditure of energy for the emission itself, as opposed to thermo- or photoemission [3]. Electron field emission can be stimulated at much lower electric fields if the cathode used has a tip of radius $10^{-5}$ – $10^{-4}$ mm. Currently, electron field emitters are widely applied in different electronic devices [4], [5], [6], [7], [8]. Compact geometrical sizes of emission area (20-30Å), and high density of emission current are major advantages of field emitters.

## MATERIALS AND THEIR SYNTHESIS

The morphology of a Ge/Si heterosystem is of great interest to nanotechnologists and physicists due to the fact that germanium islands with a size of 10-30 nm, being defect-free when embedded in silicon crystal, are quantum dots, i.e., objects in which the energy spectrum of mobile current carriers is quantized discretely along three coordinate axes. The possibility of obtaining ensembles of quantum dots by direct epitaxial synthesis is extremely attractive to nanoscientists because of its relative low cost and high performance.

To study field-emission current density, Ge/Si samples were grown by molecular beam epitaxy. Ge was first deposited on Si(100) n-type substrate doped with As ($10^{19}$ $cm^{-3}$). The epitaxial growth of Ge on a Si substrate proceeded in several stages. Two to three monolayers of germanium atoms were evenly deposited over the surface of the substrate. It is known that as growth of Ge continues on the surface of this film, many islands of pure germanium are formed, the shape, size and number of which mainly depends on the temperature and growth rate. The morphology of these germanium islands on the silicon substrate surface is presented on Fig. 1.

To protect the sample from potential contaminants, the surface of the sample was covered with a layer of antimony, which was removed by vacuum (750 torr) prior to the samples being measured. A second sample of Ge/Si (prepared as described above) was exposed to room air in the absence of Sb layer.

The purpose of this work was to determine the density of the field emission current from the surfaces of Ge/Si nanostructures to establish their suitability as multipoint field emitters. Thus, the wetting layer of Ge/Si (the Si surface covered with monolayers of Ge atoms) with additional adsorption layers (due to exposure to atmosphere) of one sample was compared with the Sb-coated (“clean”) wetting Ge/Si layer of another sample. The Ge clusters of the nanostructures with adsorption layers were compared with "pure" Ge clusters, and it was observed that the silicon content as well as adsorption layers in nanoclusters affects their size and shape.

## CURRENT-VOLTAGE CHARACTERISTICS

Scanning Tunneling Microscopy (STM) was used to measure the density of emission current from Ge/Si because of the locality of measurements and low potential difference provided by using this method of measurement. At the measurements of current-voltage characteristics between the STM probe and the Si substrate, saw-tooth voltage was applied. Complete movement of the probe (in the limits of the selected area) was carried serially, and at each step the dependence of emission current $I$ on voltage $U$ was measured. The potential of the surface was maintained at 0 V. The results of the measurements of current-voltage characteristics of Ge/Si nanostructures are presented in Fig 2.

In the experiment conducted, the mechanism of electron emission from the surface of Ge/Si is the tunneling of electrons though the potential barrier between a surface of the sample studied and the STM tip. The dimension and the shape of the potential barrier, as well as the absolute value of the vacuum gap, however, were unknown parameters.

To interpret the experimental data obtained for $I(U)$, it was necessary to determine to what extent the approximation of the one-dimensionality of the potential barrier is legitimate, which is one of the main approximations of the phenomenological model of the field emission mechanism, known as the Fowler-Nordheim theory. According to this theory, for a one-dimensional potential barrier, the dependence of emission current on applied voltage in coordinates $(U, I)$ is an exponential function (Fig.2), and in coordinates $(U^{-1}, \log(I/U^2))$ it is a linear function. The experimental data $I(U)$, therefore, were approximated with a linear function in $(U^{-1}, \log(I/U^2))$ coordinates (Fig. 3).

## EMISSION CURRENT DENSITY

The data obtained (Fig. 2 and Fig. 3) demonstrate that in the corresponding coordinates, the experimental curves can be satisfactorily approximated either by an exponential or by a straight line that allowed for the use of the Fowler-Nordheim equation (1) to evaluate the density of the emission current, $J$.

$$J = 1.537 \times 10^{-6} \times \frac{F^2}{\phi \times t^2(y)} \times exp\left[-6.83 \times 10^7 \times \frac{\phi^{3/2}}{F} \times v(y)\right] \quad (1),$$

where $t(y)$ and $v(y)$ are Nordheim tabulated functions [9], [10] of the argument

$$y = 3.79 \times \frac{\sqrt{F}}{\phi}, \quad (2)$$

where $F$ is an applied electric field, and $\phi$ is the work function of electrons.

The function $t(y)$ in the preexponential factor is close to unity and changes slightly with the argument $y$. The function ν(y) takes into account the influence of the image forces on the triangular potential barrier.

The Fowler - Nordheim law [11] is generally represented by the following expression:

$$J = \frac{A \times F^2}{\phi} \times \exp\,[-\frac{B \times \phi^{\frac{3}{2}}}{F} \times v(y)], \quad (3)$$

where $A$ and $B$ are constants.

The argument $y$ of the Nordheim function represents the ratio of the Schottky work function reduction to the work function at the absence of an electric field, $F = 0$. The argument $y < 1$, since the equality $y = 1$, corresponds to the complete removal of the Schottky barrier. With a satisfactory approximation (the error does not exceed approximately 0.1%), the Nordheim function for mid-range values of $F$ can be represented as follows:

$$v(y) = 0.965 - 0.739y^2. \quad (4)$$

Substituting (4), as well as the values of the constants A and B in (3), we arrive at expression (1) for field electron emission current density.

Formula (3) is accurate assuming that electrons in a metal obey the Fermi-Dirac statistics:

$$f(E) = [1 + \exp\left(\frac{E - E_f}{kT}\right)]^{-1}$$

The Fowler - Nordheim theory is strictly applicable only for $T = 0$. However, even at $T > 0$, as long as $kT < \phi$, the thermal excitation of the electrons only slightly smears the Fermi boundary in the range of several $kT$, and formula (3) retains its value. The temperature range in which the Fowler - Nordheim theory is fulfilled depends on the work function of the emitter. With an increase of $T$ and a decrease of $F$, field electron emission becomes thermionic in nature, enhanced by the field (Schottky effect).

The transparency of the barrier was calculated using either a semiclassical approximation or by the Wenzel, Kramer, and Brillouin method (WKB). The qualitative condition of the semiclassical theory used implies that the De Broglie wavelength of the particles is small in comparison with the characteristic dimensions $L$, which determine the conditions of this particular situation. This condition indicates that the wavelength of the particles should change little over distances of the order itself. The WKB method is applicable when the transparency of the potential barrier $D < 0.1$ [12].

The Fowler-Nordheim theory has been confirmed in numerous experiments for metal tip emitters with a change in the emission current within more than six orders of magnitude [13]. For semiconductor emitters, agreement between the Fowler-Nordheim theory and experimental data is observed in the so-called initial section of the current-voltage characteristic when the emission current changes within two to three orders of magnitude [14]. This circumstance allows one to use the Fowler - Nordheim equation (1) to calculate the emission current density from Ge/Si nanostructures.

Assuming in equation (1) $J = \frac{I}{S}$, where S is the area of the emitting surface in cm$^2$, $F = \beta \times U$, and where $\beta$ is the geometric factor or field factor in cm$^{-1}$, ( $U = 5\ V$ ), as well as taking into account the data shown in Fig. 3, the following estimations can be made for the emission process from the surface of a pure Ge nanocrystal:

$$log\left(\frac{1.537 \times 10^{-6} \times \beta \times S}{\phi \times t^2(y)}\right) \cong -7.93$$

and (5)

$$\frac{6.83 \times 10^7 \times \phi^{3/2}}{\beta} \times v(y) \cong 11.75$$

Considering $\phi \approx 1\ eV$, $t(y) \approx 1.044$, and $v(y) \approx 0.7$, this system of equations (5) was solved with the respect to $\beta$ and $S$. Then, using the values calculated, the density of the field emission current, $J$, was determined. The results obtained for the samples considered are presented in Table 1.

The density of emission current from Ge nanocrystal exceeds the density of emission current from the wetting layer of Ge/Si. The density of emission current of pure Ge nanocrystal is less than the density of emission current of Ge nanocrystal with adsorption layers. Thus, the results of the study demonstrate the effectiveness of the usage of the Ge/Si nanostructures grown as multiple-point emitters [15], [16].

## CONCLUSION

The features of field-electron emission from semiconductors are associated with several factors, among which are that the surface states of charge carriers can affect the characteristics of field electron emission, and the current-voltage characteristics reflect the band structure of semiconductors. STM was used to obtain current-voltage characteristics and to study local properties of Ge/Si. The mechanism of electron emission from the surface of Ge/Si nanostructures observed was indicated by the tunneling of electrons through a potential barrier at the interface between the surface area investigated and the tip of the probe.

To interpret the experimental data obtained, a phenomenological model of the field emission mechanism known as the Fowler-Nordheim theory was used, since in the corresponding coordinates, the experimental results can be satisfactorily approximated by either an exponential or a straight curve. The experimental data also allowed the application of the Fowler-Nordheim equation to estimate the emission current density.

The calculated density of emission current over Ge nanocrystal exceeds the density of emission current over Ge/Si wetting layers. The density of the emission current of a pure Ge nanocrystal is less than the density of the emission current of Ge nanocrystal with adsorption layers. The estimates of numerical values for field emission current density are in agreement with results of studies of field-emission phenomena from the surface of pointed semiconductor emitters [14].

## Figure and Table Captions

Figure 1. Morphological characteristics of Ge islands on the Si (100) surface

Figure 2. Current-voltage characteristics of Gi/Si nanostructures (1 – wetting layer of Ge/Si (the Si surface covered with monolayers of Ge atoms) with additional adsorption layers, 2 – Ge with adsorption layers, 3 – pure Ge, 4 – pure wetting layer of Ge/Si)

Figure 3. Current-voltage characteristics of Ge/Si nanostructures plotted in Fowler-Nordheim coordinates

Table 1. Dependence of density of emission current on voltage of 5V

Figure 1.

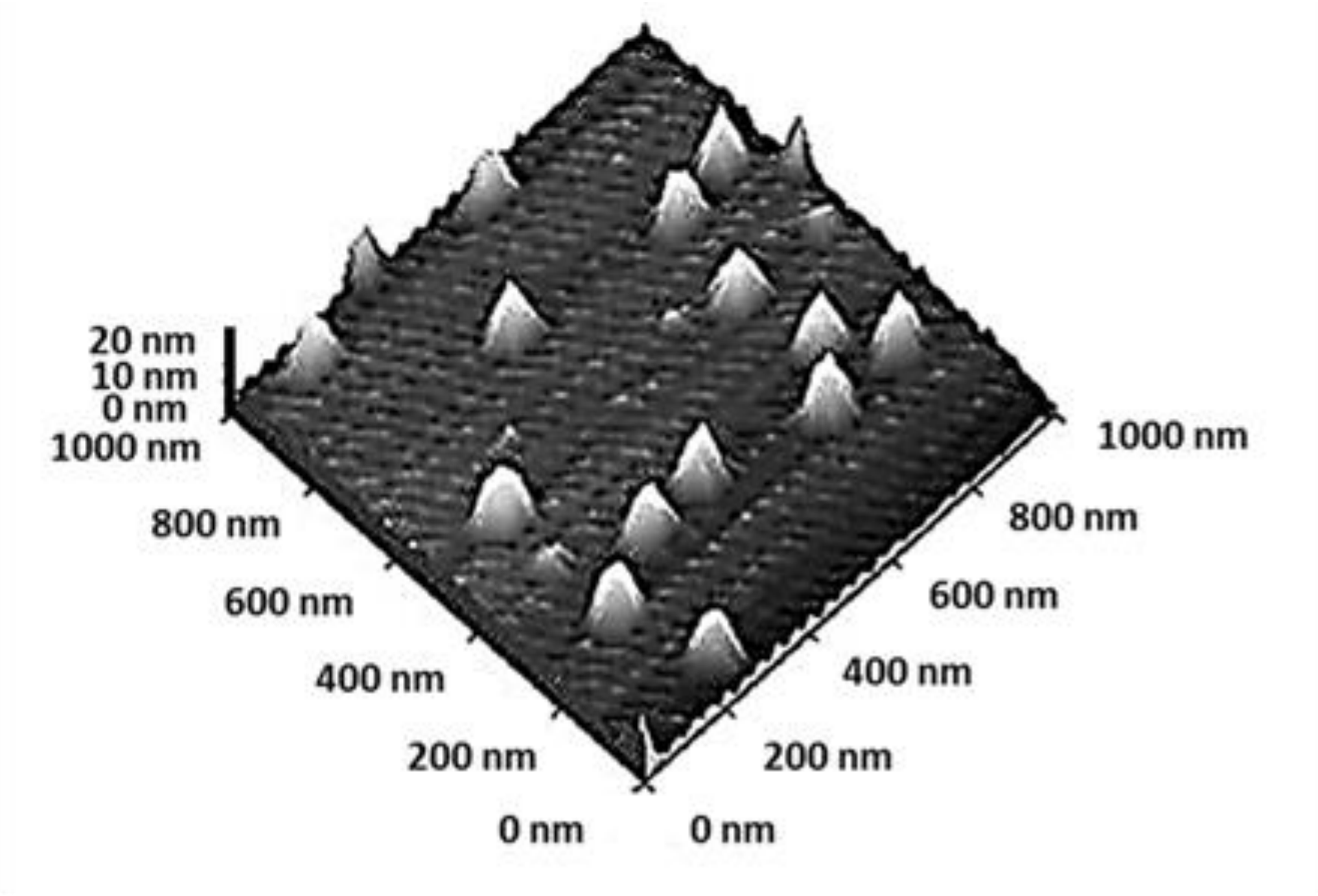

Figure 2.

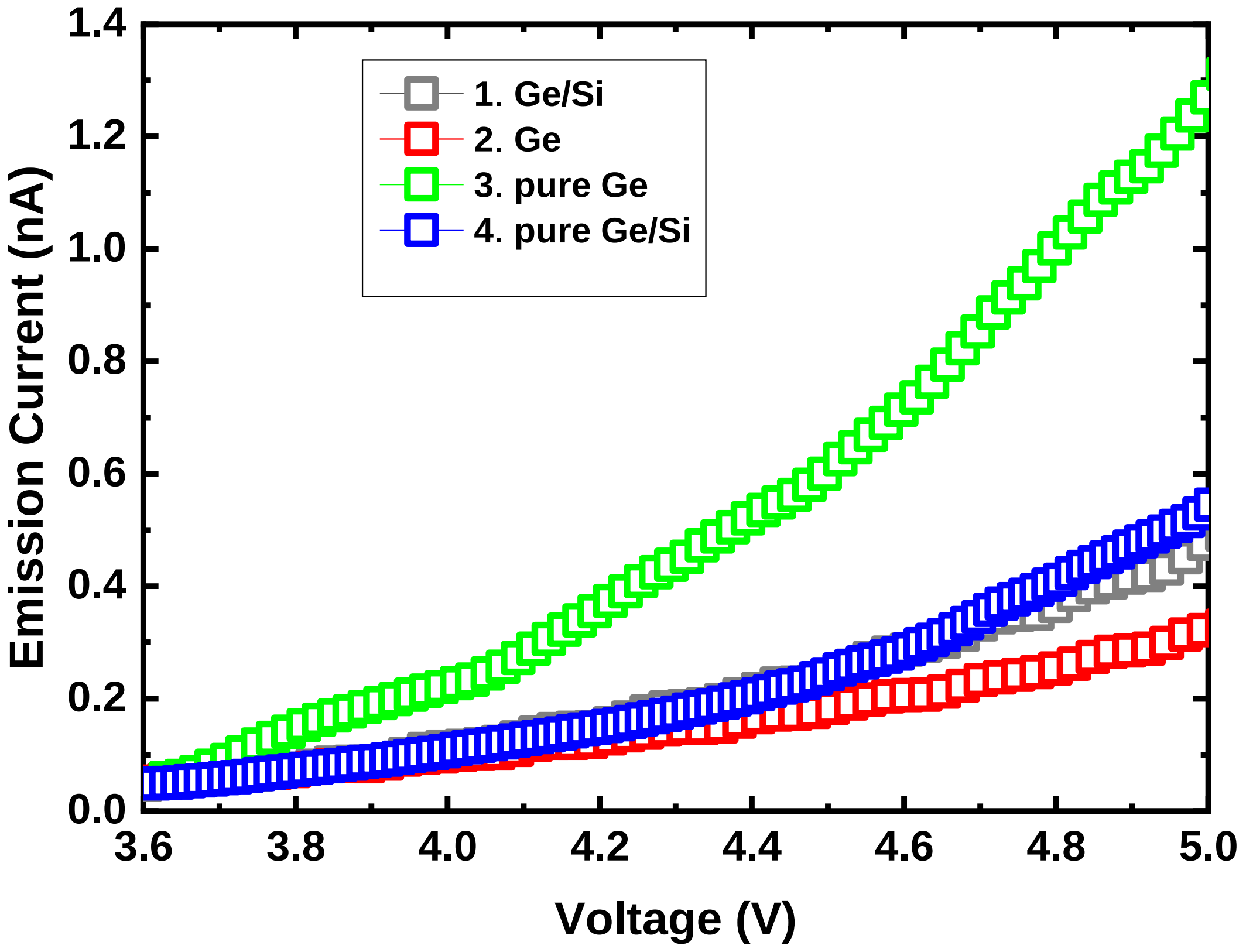

Figure 3.

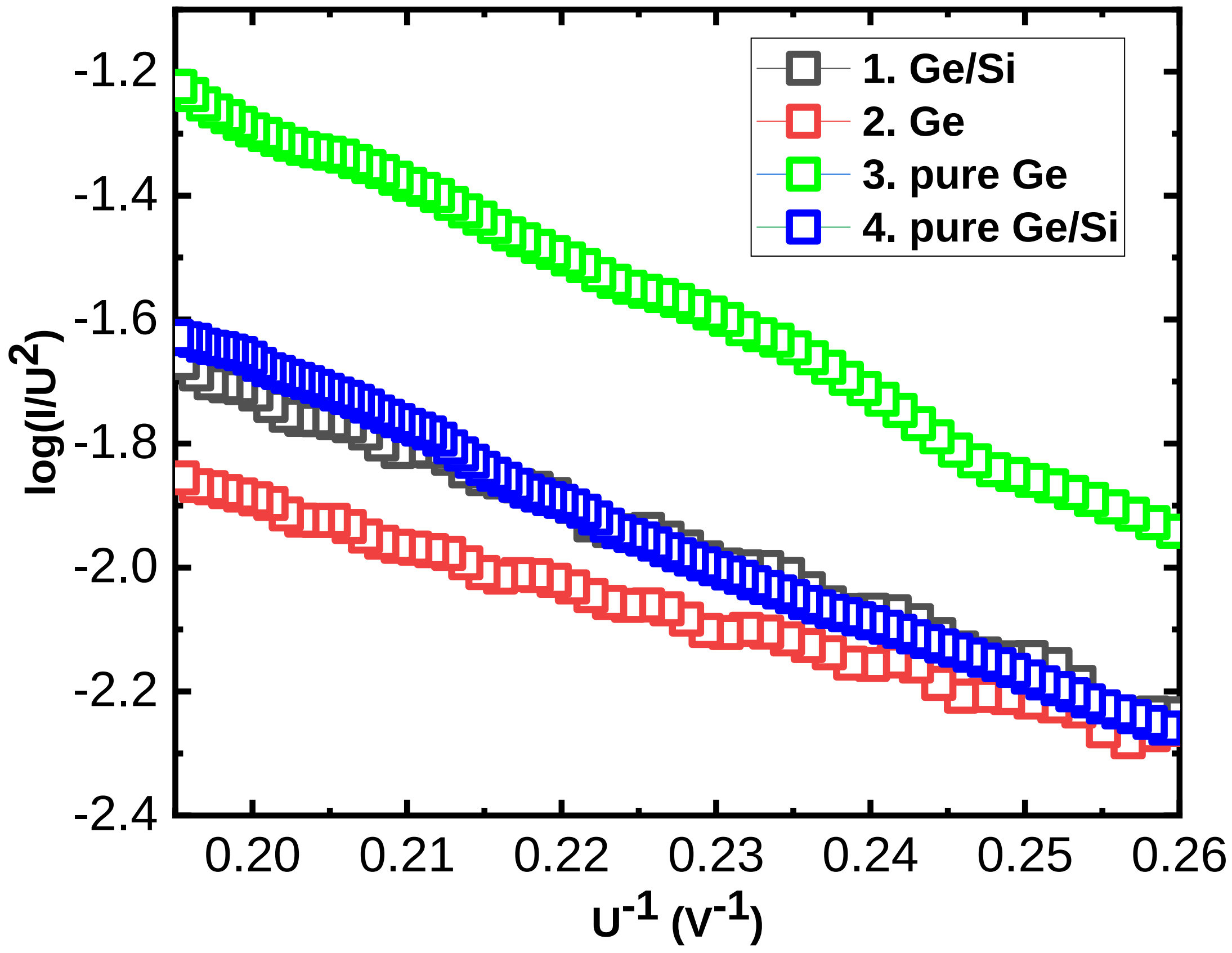

Table 1.

| | Material | $I \cdot 10^{-9}$ A | $\beta \cdot 10^{6}$ $cm^{-1}$ | $J \cdot 10^{6}$ $A/cm^{2}$ |
|---|---|---|---|---|
| 1 | Ge/Si | 0.491 | 2.280 | 2.768 |
| 2 | Ge | 0.330 | 3.118 | 15.960 |
| 3 | Pure Ge | 1.311 | 2.767 | 6.492 |
| 4 | Pure Ge/Si | 0.545 | 2.076 | 1.520 |